\documentclass{iau}
\usepackage{graphicx,natbib,amsmath,hyperref,epsfig}

\newcommand{\beq}{\begin{equation}}
\newcommand{\eeq}{\end{equation}}
\newcommand{\ds}{\ensuremath{\Delta\Sigma}}
\newcommand{\jcap}{J. Cosmology Astropart. Phys.}
\newcommand{\ssr}{Space Sci. Rev.} 
\newcommand{\apj}{ApJ}           
\newcommand{\mnras}{MNRAS}       

\newcommand{\aap}{A\&A}

\newcommand{\aj}{AJ}

\newcommand{\apjs}{ApJS}           

\newcommand{\physrep}{Phys. Rep.}

\title{Galaxy Halo Masses from Weak Gravitational Lensing}

\author[Mandelbaum]{Rachel Mandelbaum$^1$}
\affiliation{$^1$McWilliams Center for Cosmology, Department of Physics, Carnegie Mellon University, Pittsburgh, PA 15213, USA.  email: {\tt rmandelb@andrew.cmu.edu}}

\pubyear{2014}
\volume{311}
\jname{Galaxy Masses as Constraints of Formation Models}
\editors{M. Cappellari \& S. Courteau, eds.}

\begin{document}

\maketitle

\begin{abstract}
  In this review, I discuss the use of galaxy-galaxy weak lensing
  measurements to study the masses of dark matter halos in which
  galaxies reside.  After summarizing how weak gravitational lensing
  measurements can be interpreted in terms of halo mass, I review
  measurements that were used to derive the relationship between
  optical galaxy mass tracers, such as stellar mass or luminosity, and
  dark matter halo mass.  Measurements of galaxy-galaxy lensing from
  the past decade have led to increasingly tight constraints on the
  connection between dark matter halo mass and optical mass tracers,
  including both the mean relationships between these quantities and
  the intrinsic scatter between them.  I also review some of the
  factors that can complicate analysis, such as the choice of modeling
  procedure, and choices made when dividing up samples of lens
  galaxies.
\end{abstract}

\firstsection
\section{Introduction}

The field of galaxy formation and evolution seeks to explain the
evolutionary history of galaxies, but is handicapped by the
difficulties in observing the dark matter field in which the galaxies
form and evolve.  The connection between galaxies and their host dark
matter halos is an essential ingredient in the physics of galaxy
formation.  One very useful probe of the galaxy-dark matter connection
is weak lensing around galaxies, or galaxy-galaxy lensing
\citep[e.g.,][]{2008ARNPS..58...99H}.

Gravitational lensing, the deflection of light by mass, induces
tangential shear distortions in the shapes of background galaxies
around foreground galaxies, allowing direct measurement of the
galaxy-matter correlation function.  This approach has the advantage
of being sensitive to {\em all} matter, independent of its dynamical
state.  The individual distortions are small (typically of order
0.1\%), but by averaging over all foreground (``lens'') galaxies
within a given subsample selected based on their properties (a process
known as ``stacking''), and over all the source galaxies behind them,
it is possible to obtain a high signal-to-noise measurement of the
shear as a function of angular separation from the galaxy, known as
the galaxy-galaxy lensing signal.

In the past $\sim 15$ years, the quantity of imaging data with
high-quality galaxy shape information and with some redshift
information (either spectroscopic or photometric redshifts) has
drastically increased, in large part due to the efforts of major
surveys such as the Red-Sequence Cluster Survey
\citep[RCS,][]{2005ApJS..157....1G} and RCS2
\citep{2011AJ....141...94G}, the Sloan Digital Sky Survey
\citep[SDSS,][]{2000AJ....120.1579Y}, the Canada-France-Hawaii
Telescope Lensing Survey
\citep[CFHTLenS,][]{2012MNRAS.427..146H,2013MNRAS.433.2545E}, and the
{\em Hubble Space Telescope} ({\em HST}) COSMOS Survey
\citep{2007ApJS..172..196K,2007ApJS..172...38S,2007ApJS..172....1S}.
The increase in data quantity and quality, plus the appeal of a
measurement method that is directly sensitive to the dark matter, has
led to tremendous progress in the use of galaxy-galaxy lensing to
study the connection between galaxies and matter.

The purpose of this review is to describe some of the studies that
have analyzed galaxy-galaxy weak lensing measurements to learn about
the masses of dark matter halos around galaxies, and to highlight
important lessons learned and unsolved aspects of this problem that
require more work in future, for example for the next generation of
large lensing surveys that will measure weak gravitational lensing
more precisely than has been done with current datasets.  The topic of
this review is limited to relatively general studies of typical
galaxies, leaving aside studies of particular galaxy types using weak
gravitational lensing
\citep[e.g.,][]{2009MNRAS.393..377M,2010MNRAS.407....2D,2012MNRAS.425.2610R},
studies of the radial profiles of dark matter halos with weak lensing
\citep[e.g.,][]{2007ApJ...667..176G,2008JCAP...08..006M,2010MNRAS.408.1463S},
and studies of group or galaxy cluster dark matter halos
\citep[e.g.,][]{2013SSRv..177...75H}.

\section{Theory}

Weak gravitational lensing, the deflection of light rays by the mass
in matter along the line-of-sight, results in coherent shape
distortions (shears) in the background galaxies.  The strength of the
lensing shear depends on the mass in the lens object (galaxy or
cluster), the separation on the sky between the lens and the source
object, and the line-of-sight distances to lens and source.

A galaxy-galaxy weak lensing measurement probes the connection between
galaxies and matter via the cross-correlation functions
$\xi_\text{gm}(\vec{r})$, which can be related to the projected surface
density of matter around the lens galaxies
\beq\label{E:sigmar}
\Sigma(R) = \overline{\rho} \int \left[1+\xi_\text{gm}\left(\sqrt{R^2 + \chi^2}\right)\right] d\chi.
\eeq
where $R$ is the transverse separation and $\chi$ the radial direction
over which we are projecting\footnote{We are ignoring the effects from the 
radial window, which is broad enough that it is not relevant at galaxy
scales; see section 2.3 of \cite{2001MNRAS.321..439G} for
details.}. The surface density is then related to the observable 
quantity for lensing, the differential surface density, 
\beq\label{E:ds}
\ds(R) = \gamma_t(R) \Sigma_c= \overline{\Sigma}(<R) - \Sigma(R), 
\eeq
in the weak lensing limit, for
a matter distribution that 
is axisymmetric along the line of sight (which is naturally achieved
when stacking thousands of lens galaxies to determine their
average lensing signal).  Typical galaxy-galaxy lensing measurements
require substantial numbers of lens and source galaxies in order to
average out the random component of the source galaxy
shapes, which is typically the dominant source of noise in the
measurement.
This observable quantity can
be expressed as the product of two factors, a tangential shear
$\gamma_t$ and a geometric factor
\beq\label{E:sigmacrit}
\Sigma_c = \frac{c^2}{4\pi G} \frac{D_S}{D_L D_{LS}}
\eeq
where $D_L$ and $D_S$ are angular diameter distances to the lens and
source, and $D_{LS}$ is the angular diameter distance between the lens
and source. 

For isolated galaxies at the center of dark matter halos, the
surface density of surrounding matter can be written in terms of the 3d
density profile $\rho(r)$ on small scales, as 
\beq\label{E:sigmar2}
\Sigma(R) = \int_{-\infty}^{\infty} \rho(r=\sqrt{\chi^2+R^2}) d\chi.
\eeq
However, there are additional contributions due to those galaxies that
are satellites, for which the host dark matter halo leads to a
contribution for $R$ in the hundreds of kiloparsec to 1 Mpc range.  On
even larger scales, there is a small but significant contribution due
to the matter in structures that are physically associated with but
not inside of the halo in which the lens galaxy resides (the ``2-halo
term'').  As a result, the interpretation of the lensing signal
can be complicated, depending on the local environment of the galaxies
(mixture of isolated and central galaxies vs. satellites) and the
separations $R$ used (larger scales requires modeling of the term due to
large-scale structure).  Moreover, the stacking process means that the
average profile can be affected not just by mean relationships between
mass and observables, but also by scatter in those relationships.
Different studies have taken different approaches to these
interpretation challenges, as I will describe below.

\section{Results}\label{sec:results}

Among the first galaxy-galaxy lensing results with detailed modeling of the 
mass distributions of large samples of foreground lens galaxies were
\cite{2005ApJ...635...73H}, \cite{2006MNRAS.368..715M}, and
\cite{2006MNRAS.371L..60H}.

\cite{2005ApJ...635...73H} used data from the RCS to identify isolated
galaxies and measure their lensing signals as a function of stellar
mass or luminosity.  Due to the use of isolated galaxies, their
approach was to fit the signals to an NFW \citep{1996ApJ...462..563N}
profile, ignoring the possibility of contributions from host halos of
those galaxies that may be satellites in larger halos, and restricting
to small enough scales that the 2-halo term is 
negligible.  They find a relationship
between halo mass and luminosity that goes like $M_\text{halo}\propto
L^{1.5}$. This result is based on the best-fit mass for samples with
some average luminosity, ignoring scatter between mass and luminosity.
They also made these measurements for early and late-type galaxy
samples split based on color, and found $M_\text{halo}/M_*$ higher by
a factor of $\sim 2$ for early types.  Dividing $M_*/M_\text{halo}$ by
the cosmological baryon fraction to estimate an efficiency of
conversion of baryons to stars, 
\beq\label{eq:eta}
\eta = \frac{M_*}{M_\text{halo}} \frac{\Omega_\text{m}}{\Omega_\text{b}}
\eeq
they found $\eta\sim 33$\%
and $\sim 14$\%, respectively, for early and late type galaxies.

Using data from the SDSS Main spectroscopic galaxy sample,
\cite{2006MNRAS.368..715M} analyzed the host halo mass for early and
late-type central galaxies as a function of their stellar mass and
luminosity, for galaxies at a typical redshift of $\langle z\rangle\sim 0.1$ (lower
than the RCS study).  In this work, the split
into early versus late types was achieved using a morphological
estimator (not color).  This work also estimated the satellite
fractions purely based on the lensing signal alone.  In order to
interpret the lensing signals in terms of central and satellite galaxies in a
statistical sense, this work used a halo model
\citep[e.g.,][]{2000MNRAS.318..203S,2002PhR...372....1C} formalism
that was tested for these purposes on mock galaxy samples derived from
$N$-body simulations by \cite{2005MNRAS.362.1451M}.  The simple halo
model that was used had only two free parameters, with the rest fixed
to some values selected based on the $N$-body simulation analysis, and
correction factors for scatter in the mass-observable relation were
applied to the best-fit masses.  The findings for 
$M_*/M_\text{halo}$ in this work were consistent with those from RCS
for stellar masses above $10^{10}M_\odot$.  Below that stellar mass,
early and late type galaxies were found to have statistically
consistent conversion efficiencies (Eq.~\ref{eq:eta}), suggesting that stellar mass is a
good tracer of halo mass below $M_*\sim 10^{10}M_\odot$.  For a galaxy sample
around $L_*$, the halo model analysis of the lensing results suggests
$M_\text{halo}/L=79^{+27}_{-24}$ and $41^{+16}_{-17} M_\odot/L_\odot$
($2\sigma$) for early and late types, respectively, with
higher values at higher luminosity; results at lower luminosity are
too noisy to draw conclusions about trends in that direction.

Using data from the {\em HST} GEMS survey,
\cite{2006MNRAS.371L..60H} analyzed the lensing signal from high
stellar mass galaxies over a long redshift baseline,
$0.2<z<0.8$.  Given the stellar mass limit $\log{\left(M_*/M_\odot\right)}>10.5$,
the majority of the galaxies exhibit early type morphology; however,
no explicit morphological split was used.  After modeling the signals
using a pure NFW profile without modeling of signal due to satellites
or halo vs. stellar mass scatter, they find an average conversion
efficiency of $\eta=0.10\pm 0.03$ ($1\sigma$) for the entire sample.  These
results are consistent with those from lower redshift
\citep{2005ApJ...635...73H,2006MNRAS.368..715M} when comparing with
samples that have a similar stellar mass range.  When 
splitting into redshift bins within the GEMS sample, they find no
statistically significant evolution of the $M_\text{halo}/M_*$ ratio, though
their best-fit relation includes slight evolution in the direction
of higher ratio at higher redshift, giving an upper limit in growth of
this ratio of $\lesssim 2.5$ from $z=0.8$ to the present time.

More recently, \cite{2012ApJ...744..159L} carried out a joint analysis
of the galaxy-galaxy lensing, galaxy clustering, and abundance of
galaxies as a function of stellar mass (without splitting by galaxy
type) in the COSMOS survey.  While this survey covers a very small
area of the sky, its depth provides a long redshift baseline over
which to carry out this analysis.  This joint analysis used a quite
complex formulation of the halo model with many parameters including
ones regulating the scatter between the halo and stellar mass, as
described in \cite{2011ApJ...738...45L}.  For this 
particular dataset, the stellar mass function has the highest $S/N$
and therefore dominates the constraints on the halo vs. stellar mass
relation, which is provided as a functional form with four parameters
determined in three redshift bins.  The redshift evolution of
this halo vs. stellar mass relation is indicative of downsizing, with
the stellar mass at which the star baryon conversion efficiency is
maximized decreasing at lower redshifts.  One new aspect to this halo model
analysis compared to previous ones used for lensing analyses is that
it provides constraints on the logarithmic scatter in the stellar mass at fixed
halo mass, which varied in the range $\sigma_{\log{M_*}}=0.2$--$0.25$
for the three redshift bins.

The halo model formalism used by \cite{2012ApJ...744..159L} relied on
the use of all galaxies at fixed stellar mass to define the abundance.  An
updated version of the formalism that permits a split into
star-forming and passive galaxies was presented and used by
\cite{2013ApJ...778...93T}, also on the COSMOS data.  The results of
that work suggested that for massive galaxies ($M_*\gtrsim
10^{10.6}M_\odot$), star-forming galaxies form stars at a rate that
roughly matches the growth of their dark matter halos via accretion
from $z=1$ to $0$, whereas for quiescent galaxies, the growth of their
dark matter halos outpaces star formation, reducing their apparent
baryon conversion efficiency $\eta$.  In contrast, for lower mass galaxies,
the halo vs. stellar mass relations are similar for passive and
star-forming galaxies at $z<1$.  These findings are qualitatively
consistent with the trends from the SDSS data using lensing alone in
\cite{2006MNRAS.368..715M}, but the $S/N$ in COSMOS is superior for
lower mass galaxies.  Moreover, COSMOS enables tests of redshift
evolution, and find a cross-over in the red-blue stellar vs. halo mass
relation as a function of redshift, which has not been seen in other
surveys (either due to insufficient redshift baseline, or different
modeling strategies).

\cite{2011A&A...534A..14V} used imaging data from the RCS2 together
with SDSS spectroscopic redshifts in order to perform a weak lensing
study of the halo vs. stellar mass relation, with a similar halo model
formalism as that used in \cite{2006MNRAS.368..715M}.  They find a relatively
steep scaling of halo mass with luminosity, $2.2\pm 0.1$ and $1.8\pm
0.1$ for red and blue galaxies, respectively\footnote{These numbers
  differ from those in \cite{2011A&A...534A..14V}, and
  come from a correction that was reported in
  \cite{2014MNRAS.437.2111V}, which used the same halo modeling
  software.}, which is inconsistent with that found by
\cite{2005ApJ...635...73H} in the RCS; however, the different methods
of modeling the lensing signals and different selection criteria used
could be responsible for this difference.  The results are consistent
with those from the SDSS analysis by \cite{2006MNRAS.368..715M} within
the errors.  When splitting the sample by stellar mass,
\cite{2011A&A...534A..14V} confirm the findings from the SDSS: below
$10^{11}M_\odot$, the stellar mass traces halo mass, but above that
stellar mass, the halo masses differ for early and late type galaxies.

\cite{2013MNRAS.430..725V} presented a 9-parameter conditional
luminosity function (CLF) formalism including various input from
$N$-body simulations (halo mass function, concentration-mass relation,
and others).  \cite{2014MNRAS.437..377C} showed that when fitting the
parameters of this model to describe low-redshift SDSS samples, the
resulting CLF was able to describe lensing signals for
higher-luminosity and redshift SDSS and RCS2 samples without further
adjustment.  This result is non-trivially interesting and suggests
that the 9-parameter model may be capturing the key features of the
galaxy-dark matter connection.  One interesting aspect of this model
is that it includes 
constant scatter in $\log{L}$ at fixed $\log{M_\text{halo}}$, which
implies that the scatter in $\log{M_\text{halo}}$ at fixed $\log{L}$
is not constant (given that the slope of the luminosity-mass relation
varies).  This aspect of the model is similar to the results of the
COSMOS analysis from \cite{2012ApJ...744..159L}, which described the
results in terms of a fixed scatter in $\log{M_*}$ at fixed
$M_\text{halo}$.

Two works \citep{2013arXiv1310.6784H,2014MNRAS.437.2111V} have
explored the galaxy-galaxy weak lensing signals as a function of lens
galaxy properties using CFHTLenS data.  The CFHTLenS data cover
sufficient area with enough depth that, compared to the SDSS, RCS, and
RCS2, the statistical errors in the lensing signal at lower stellar
mass and luminosity have decreased significantly\footnote{The errors
  are similar to those in COSMOS at low stellar mass, however in the
  case of the COSMOS analysis, the clustering and stellar mass
  function plays a significant role in the halo model constraints at
  low stellar mass.}.  When fitting the halo mass to a power law in
luminosity (stellar mass), \cite{2014MNRAS.437.2111V} find a power law
index of $1.32\pm 0.06$ and $1.09^{+0.20}_{-0.13}$
($1.36^{+0.06}_{-0.07}$ and $0.98^{+0.08}_{-0.07}$) for red and blue
galaxies, respectively.  This is shallower than the RCS2 results,
possibly because of multiple differences in galaxy selection, as
discussed in detail in \cite{2014MNRAS.437.2111V}; however, their
results are largely in agreement with the results from the SDSS.
Comparison with COSMOS results is somewhat more complex, given that
the best-fitting stellar vs. halo mass relationship can only be turned
into an average halo mass within each bin when considering the
scatter in stellar mass at fixed halo mass.  We will return to this
point shortly.

\cite{2014arXiv1404.6828H} carried out a maximum-likelihood weak
lensing analysis of optically-selected galaxy groups in the Galaxy and
Mass Assembly (GAMA) survey, using SDSS galaxies as background
sources.  While the focus of the paper is on groups of galaxies,
results are also shown for a set of ``groups'' with $N_\text{gal}=1$,
i.e., individual galaxies split by stellar mass (not type or
morphology).  The trends for the average halo mass as a function of
stellar mass were largely consistent with those from previous surveys.
One point that was highlighted in both \cite{2011ApJ...738...45L} and
\cite{2014arXiv1404.6828H},  and that is quite relevant
to the comparison of results from e.g. \cite{2014MNRAS.437.2111V}
against those from \cite{2012ApJ...744..159L}, is that the average
halo mass at fixed stellar mass is very sensitive to the lognormal
dispersion in $M_*$ at fixed $M_\text{halo}$ (even more so than on the
median $M_*$ at fixed $M_\text{halo}$).  Thus, comparison of results
of contraining $M_*$ as a function of $M_\text{halo}$ against
measurements of $\langle M_\text{halo}\rangle$ at fixed $M_*$ is
highly non-trivial.  For the GAMA galaxies, \cite{2014arXiv1404.6828H}
found 
$\sigma_{\log{M_*}}\sim 0.15$, consistent with the COSMOS results
within the errors.  However, since $\sigma_{\log{M_*}}$ includes not
just intrinsic scatter but also measurement error (which could depend
on how the stellar mass are estimated and whether the dataset has
spectroscopic or photometric redshifts), it is not necessarily the
case that we even expect similar results from all surveys.

\begin{figure}
\begin{center}
 \includegraphics[width=3.8in]{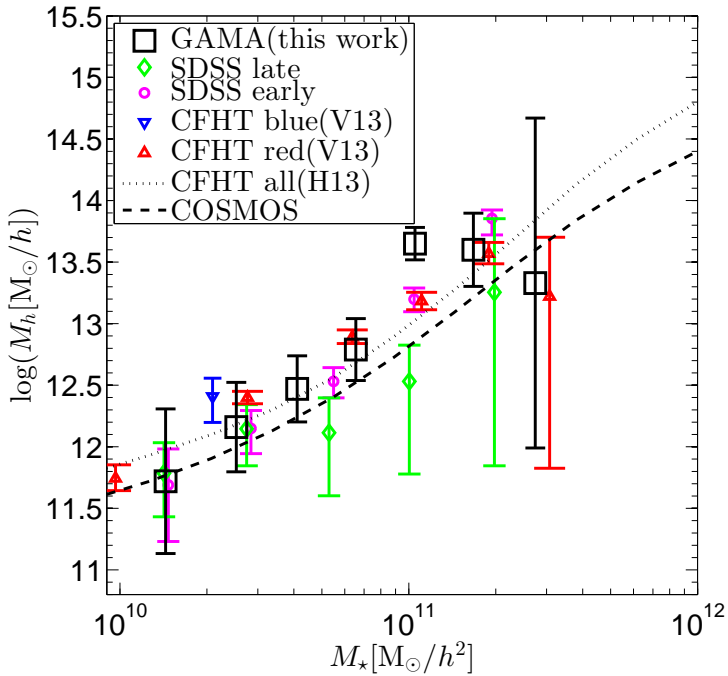}
\end{center}
 \caption{The central halo mass-stellar mass relation measured from
   weak lensing, from several works described in this section.  The quantity that is plotted is the average
   logarithmic halo mass at fixed stellar mass. The COSMOS
   measurement, originally provided as a fitted $\langle\log(M_\star)|M_h\rangle$
   relation, was converted to the plotted quantity by taking into account the
   scatter in mean stellar mass at fixed halo mass. The same
   conversion is done for the results from
   \citet{2013arXiv1310.6784H}. It is important to keep in mind that
   $\langle M_\text{halo}\rangle(M_*)$ 
   depends not just on the mean relationship between these quantities,
   but also on both their intrinsic scatter and the measurement error
   in the $M_*$.  For example, surveys with stellar masses based
   purely on photometric redshifts, which are an additional source of
   non-negligible scatter, should have lower average halo masses at
   fixed stellar mass.  Figure used with permission from
   \cite{2014arXiv1404.6828H}.}\label{fig:comp}
\end{figure}

To summarize, in Fig.~\ref{fig:comp} we show figure 9 from
\cite{2014arXiv1404.6828H}, which compares the results for the average
halo mass in stellar mass bins (averaged over all galaxy types) from
several different papers.

\section{Future challenges}\label{S:challenges}

Aside from the basic observational challenges in weak lensing
measurements, which are summarized in
e.g. \cite{2013PhR...530...87W}, here I discuss some of the challenges
in how to make and interpret weak lensing measurements of dark matter
halo masses.

The first issue is in defining the observational quantity to be
used as a mass tracer, typically either stellar mass and luminosity.
For galaxy samples with spectra, the presence of a redshift can be
useful in making a robust estimate of these quantities by fitting the
SED to templates.  However, for lens samples with imaging data alone,
photometric redshifts uncertainty becomes uncertainty in the
stellar mass and luminosity.  The process of carrying out these fits
to interpret the SED in terms of stellar mass and luminosity has other
sources of uncertainty.  The most commonly considered uncertainty is
due to the stellar initial mass function (IMF), which can
lead to tens of percent differences in stellar mass estimates.
However, even controlling for IMF differences,
\cite{2012ApJ...746...95L} argue that differences in stellar mass
estimates due to the stellar population synthesis model, form chosen
for star formation history, and ways of handling dust attenuation can
lead to $\sim 45$\% uncertainties in the stellar mass estimates.
Finally, in the estimate of stellar mass from the mass-to-light ratio
and luminosity, we cannot ignore uncertainties that may arise due to
the flux estimate that goes into luminosity.  This estimate often
involves fitting the surface brightness profile of the galaxies, which
itself has many pitfalls in the choice of models to use
\citep[e.g.,][]{2014MNRAS.443..874B} and systematic errors due to sky
level misestimates near bright foregrounds.  These sources of
uncertainty make comparison between measurements from different
surveys with different analysis techniques necessarily difficult, and
in future, we will need to control for such differences quite
carefully.

Another issue, which is relevant for studies that attempt to define
type-dependent relationships between stellar and halo mass, relates to
how the galaxies were split into different types.  Morphological
estimators can be noisy and difficult to apply to limited-resolution
imaging from the ground.  Estimators based on spectra, such as the
strength of the $4000$\AA\ break, are limited to the special cases
where spectra for lens objects are available.  Use of rest-frame
colors from imaging data can couple photometric redshift errors to the
quantity used for type classification, and can also be fooled by
dust extinction in edge-on spirals.  It is therefore not
apparent to what degree we can compare type-dependent relationships
that are derived in different ways.  It is interesting that
\cite{2013arXiv1310.6784H} and \cite{2014MNRAS.437.2111V} found using
CFHTLenS that for certain stellar mass bins, their estimates of the
halo vs. stellar mass relation differ for certain galaxy types.
Naively, since they are using the same source catalog and photometric
redshifts, one would expect the noise in their measurements to be
correlated, which should result in their agreeing at better than the
$1\sigma$ level. 
However, the two works differ in their choice of type separator (color
vs. spectral type, respectively) and their way of binning the galaxies
by stellar mass (directly vs. using luminosity and then getting an
average stellar mass for each luminosity bin).  They also have some
differences in how the mass modeling of the lensing signals was
carried out.  Differences in how the data were split into samples and
modeled must be responsible for the difference between these results.

The final challenges have to do with how halo mass is modeled.  The
choice of whether to use isolated galaxies (which do not require
halo modeling) or not could lead to the selection of unfair
samples compared to use of all galaxies regardless of
environment.  At the same time, the halo model that can be used to
interpret the lensing signal for a sample of mixed central and
satellite galaxies has its own
limitations: for example, it does not include assembly bias (wherein galaxy
properties depend not just on halo mass but some other parameter like
age; see \citealt{2014MNRAS.443.3044Z} regarding its effects on
halo model analyses, and \citealt{2014MNRAS.444..729H} for an analysis
that attempts to account for assembly bias).  It also relies on $N$-body based quantities like
the halo mass function and concentration vs. mass relation, which
means that if baryonic effects are important
\citep{2010MNRAS.405.2161D,2014MNRAS.441.1769C} then the halo model
predictions will be incorrect and therefore so will the estimated
masses.  Another factor that is relevant to the studies that already
use halo models is the choice of how much complexity to allow.  For
example, \cite{2006MNRAS.368..715M}, \cite{2011A&A...534A..14V}, and
\cite{2014MNRAS.437.2111V} all used simple halo models with many
parameters fixed and with simple corrections for halo mass
vs. observable scatter; \cite{2011ApJ...738...45L} and
\cite{2013ApJ...778...93T} used a much more complex halo model with
more freedom and fewer fixed parameters, but they also needed to
include much more data (stellar mass function and clustering) in order
to get strong parameter constraints, which means that systematic
uncertainties in abundances become more relevant than in the
lensing-only studies.  It is clear that some of these nuisance
parameters are important, and in order to compare results we must
understand the impact of choices about which nuisance parameters are held fixed
and which are free.

In short, the above complications can complicate a comparison between the
different measurements described in Section~\ref{sec:results}.
 In the near future, larger lensing surveys are going to
make even higher $S/N$ measurements of galaxy-galaxy lensing, which
could lead to 
a better understanding of the halo vs. dark matter connection from
weak lensing.  At the same time, with the better data will come a need
for greater understanding of how these challenges are affecting the
results.

\section{Conclusions}

Because of its sensitivity to all types of matter (stars, gas, and
dark matter) and its insensitivity to their dynamical state, weak
gravitational lensing has emerged as a powerful tool to study the
connection between galaxies and their host dark matter halos in the
past decade, with results from multiple surveys, the most recent of
which have achieved statistical errors in the 5\% regime.  While there
are a number of challenges to understand in the modeling that is used
to go from measured lensing signals to constraints on masses, as
described in \S\ref{S:challenges}, there is every reason to believe
that near-term surveys like Hyper Suprime-Cam
(HSC\footnote{\url{http://www.naoj.org/Projects/HSC/index.html}},
\citealt{2006SPIE.6269E...9M}), the Dark Energy Survey
(DES\footnote{\url{https://www.darkenergysurvey.org/}},
\citealt{2005astro.ph.10346T}), and the KIlo-Degree Survey
(KIDS\footnote{\url{http://www.astro-wise.org/projects/KIDS/}}) will
continue to improve on what we have already learned, and we will learn
yet more from the even more ambitious programs that are planned for
the coming decade, including the Large Synoptic Survey Telescope
(LSST\footnote{\url{http://www.lsst.org/lsst}},
\citealt{2009arXiv0912.0201L}),
Euclid\footnote{\url{http://sci.esa.int/euclid},
  \url{http://www.euclid-ec.org}} \citep{2011arXiv1110.3193L}, and the
Wide-Field Infrared Survey Telescope (WFIRST-AFTA
project\footnote{\url{http://wfirst.gsfc.nasa.gov/}},
\citealt{2013arXiv1305.5422S}).

\section*{Acknowledgements}

I am grateful for the support of the Alfred P. Sloan Foundation, which
supported my work on this review.  I would like to thank Jiaxin Han
for permitting me to use figure 9 from his paper, and  Alexie Leauthaud and Ying Zu for useful discussions.

\end{document}